\documentclass[review]{elsarticle}

\usepackage{hyperref}

\usepackage{subfig}

\usepackage{graphicx}

\usepackage{amssymb}

\usepackage{amsmath}

\usepackage[capitalise]{cleveref}

\usepackage{breqn}

\usepackage{color,soul}

\usepackage{siunitx} 

\usepackage [ngerman, english] {babel}

\usepackage{array} 

\usepackage{tabularx}

\usepackage{enumitem}

\usepackage{appendix}

\usepackage[linesnumbered,ruled, algosection]{algorithm2e}

\usepackage{xcolor}

\newcommand{\grad}{^\circ}

\journal{Computational Materials Science}

\begin{document}

\begin{frontmatter}

\title{Grain-resolved kinetics and rotation during grain growth of nanocrystalline Aluminium by molecular dynamics}


\author[mymainaddress]{Paul W. Hoffrogge}


\author[mymainaddress]{Luis A. Barrales-Mora\corref{mycorrespondingauthor}}
\cortext[mycorrespondingauthor]{Corresponding author}
\ead{barrales@imm.rwth-aachen.de}
\address[mymainaddress]{Institute of Physical Metallurgy and Metal Physics, RWTH Aachen University, D-52056 Aachen, Germany}

\begin{abstract}
Grain growth in nanocrystalline Al was studied by means of molecular dynamics simulations. The novelty of this study results from the utilization of an algorithm to resolve per-grain kinetics and orientation change from molecular dynamics data sets. To this aim, a highly efficient algorithm for the identification and reconstruction of crystallites from molecular dynamics data sets of FCC materials was developed. This method is capable of calculating specific attributes of grains, namely, volume, center of mass, average orientation and orientation spread. In addition, it provides a mapping method to track grains during time-row data sets. In the present contribution, we describe and validate the algorithm, which is then used to analyze grain growth in polycrystalline Al with a weak texture. For the conditions tested, the algorithm was able to find all of the input orientations and reconstruct the grains according to their crystallographic orientation. With the help of the developed algorithms, we studied grain growth kinetics and grain rotation. The results of the simulations showed slightly slowed-down kinetics in particular in the initial stages of grain growth and marginal rotation of the grains.
\end{abstract}

\begin{keyword}
Molecular-Dynamics \sep Grain Reconstruction \sep Post-processing \sep Parallelization \sep Grain Growth \sep Grain Rotation 
\end{keyword}

\end{frontmatter}


\section{Introduction}
\label{introduction}

\subsection{Grain growth in nanocrystalline polycrystals}

Nanomaterials have become an important source for the development of new technologies due to their extraordinary properties that allow their application in fields where conventional materials fail. Among other important technological fields, nanomaterials have been utilized in medicine (drug delivery), electronics (circuit miniaturization), aerospace industry (improved fatigue-resistant materials) and many other applications. Nevertheless, due to the high density of crystal defects per unit volume in polycrystalline nanomaterials, they are in such a state of thermodynamic non-equilibrium that provides, even at room temperature, a tremendous driving force for the elimination of the crystal defects and thus, for the transition from nano- to micro-sized grained materials. Hence, the thermal stability of nanomaterials is of utmost importance for their design and development.

The thermal stability of nanocrystalline materials is still a scientific conundrum owing to the intrinsic difficulties of investigating opaque materials with nanoscaled grains and even smaller microstructural features. Nanomaterials are prone to microstructural change because they own a very high free energy $\Delta G$ stemming from their large density of crystal defects in particular grain boundaries. Nevertheless, a polycrystal is not exclusively composed of grain boundaries. The junctions that connect grain boundaries, namely triple lines and quadruple junctions, also form part of the polycrystalline structure and affect the microstructural evolution under certain conditions. For mesoscopic grain sizes, the volume occupied by grain boundaries exceeds by orders of magnitude that of other structural elements. However, this difference becomes insignificant in nano-sized materials as triple lines, being the most frequent topological element, can occupy as much volume as grain boundaries in a microstructure. The contribution of these components to the free energy is so high that most rapid grain growth is expected and also usually observed \cite{Paul2011, Ames2008, Krill2001}.

Different approaches have been suggested for the stabilization of the microstructure in nanomaterials \cite{Saber2015,Andrievski_2013,Simoes2010}. There are basically two approaches to achieve thermal stability in nanocrystalline materials, namely, the \textit{kinetic} and the \textit{thermodynamic approach} \cite{Andrievski_2013}. In the former, it is sought to hinder  grain boundaries by incorporating obstacles for their migration \cite{BarralesMora2012,BarralesMora2008a,BarralesMora2007,Barrales-Mora2013,Barrales-Mora2012,Barrales-Mora2012a,Czubayko1998,Gottstein2005,Gottstein2002,Gottstein2010,Zhao2011a,Zhao2010,Gottstein2006,Gottstein2000,Novikov2005,Novikov2010,Novikov2016,Klinger2008}. In this context,  there is also available literature that seems to suggest that intrinsic phenomena to the grain boundary structure might aid thermal stability during the normal development of the microstructure\cite{Bernstein_2008, Upmanyu01, Holm2010, Straumal2010, BarralesMora2016}. The second approach relies on the decrease of the free energy caused by the segregation of solute atoms to grain boundaries \cite{Trelewicz2009,Chookajorn2012,Murdoch2013,Saber2013}.

\subsection{\hl{Simulation of grain growth in nanocrystalline polycrystals}}

\hl{Different computer models have been utilized in the past to simulate grain growth in nanocrystalline materials. Most of these simulations have been performed by mesoscopic models} \cite{Zoellner2012,DarvishiKamachali2012,Kamachali2010,BarralesMora2010,BarralesMora2012,BarralesMora2008a,BarralesMora2007,Diss,Barrales-Mora2013,Barrales-Mora2012,Barrales-Mora2012a,Goins2016,Anderson1984,Rollett1989}\hl{. These models rely on the properties of the structural elements of a microstructure, namely, the grain boundaries, the triple lines and the quadruple junctions. Any effect acting on these elements can be separately incorporated to the models. The advantage of these models is that they allow the simulation of relatively large representative volume elements (RVE) that can be composed of thousand to million grains and thus, the RVEs are statistically representative. The disadvantage of mesoscopic models is that the effects and physical mechanisms are pre-defined and pre-implemented. This is useful to test theories by isolating particular effects that are expected to affect considerably the evolution of a system but for obvious reasons, it is impossible to resolve the mechanisms of phenomena acting at the atomic scale. Nevertheless, a combination with experiments or atomistic simulation is most useful to complement the models and/or understand particular aspects of the experiments.

By contrast, atomistic simulations allow the determination of the mechanisms of the phenomena involved. In the case, of molecular dynamics simulations empirical interatomic potentials are utilized to describe interatomic interactions so far with great success. The disadvantages of atomistic simulations is that the volume that can be simulated in a reasonable time is very limited compared to mesoscopic simulations. This results occasionally in insufficient statistical representativeness and undesired effects of the size of the simulated volume. }Regarding grain growth, molecular dynamics have been already employed to simulate grain growth. In particular, the seminal work by Farkas et al. \cite{Farkas2006,Swygenhoven2000,Farkas2008,Monk2007,Farkas2007,Farkas2007a}, Haslam and Yamakov \cite{Haslam02,Haslam04,Yamakov2006}, Holm and Foiles \cite{Holm2010} and Srolovitz et al. \cite{Thomas2016,Upmanyu01,Zhang01,Zhang02,Zhang2006} have substantially advanced our knowledge on the mechanisms of grain growth in nanocrystalline materials. The purpose of the present manuscript is to identify particular mechanisms of grain growth in nanocrystalline materials. For this purpose, we utilized molecular dynamics simulations and developed novel characterization methods for the evaluation of computational microstructures from molecular dynamics data sets.

\subsection{Post-processing of molecular dynamics}

In molecular dynamics (MD) simulations, the analysis and interpretation of the data remain a substantial and difficult part. Certain tools are required in order to reduce the amount of information and obtain useful quantities. In several situations, information on the evolution of 3D features is necessary, for example, grains in polycrystalline aggregates. This information is, however, not delivered by most of the available methods for the characterization of MD data sets. This task remains difficult because it involves identifying crystallographically the environment of the atoms to assign an orientation respective to a coordinate system and to reconstruct from this information the volume with identical orientation i.e., the grains. Due to the mathematical complexity of rotations and reconstruction algorithms, these calculations are slow and costly for their execution during runtime and even during post-processing if large data sets are involved. Nevertheless, several methods which enable the identification of crystal structures from atomic-based data have been already developed. The most straightforward way to distinguish between some of these is to compute the coordination number, i.e. the number of nearest neighbors (NNs) or next-nearest neighbors (NNNs) of the atoms. While this is a fast tool and also easy to implement, it obviously cannot distinguish between different atomic structures exhibiting the same coordination number. Another technique which allows a much more reliable identification of crystalline structures was first introduced by Honeycutt and Andersen \cite{Honeycutt}. This method analyzes the extent and connectivity properties of atom-diagrams which are comprised of nearest neighbors to two adjacent atoms. Based on this principle, also known as Common Neighbor Analysis or CNA, several algorithms have been implemented \cite{Stukowski01, Stukowski02, Begau} and successfully performed to MD simulation data. Similarly, the centro-symmetry parameter,  energy filters, bond order and angle and Voronoi analysis can also be used to discriminate atoms arranged in specific crystal structures \cite{Stukowski01}. Nevertheless, none of these methods can distinguish between groups of atoms with the same crystallographic orientation. To differentiate between differently oriented crystals, the definition of order parameters has been utilized. The drawbacks of this approach are that the orientations ought to be known beforehand and that complex rotations of the crystals might out-range the scope of the parameter. Nevertheless, this approach has been successfully utilized in numerous simulations to track, for instance, grain boundaries \cite{Janssens, Zhang01, Zhang02, Zhang03, Zhang04} and grain rotation \cite{Mishin01, Mishin02, Trautt01, Trautt02, Barr01}.
\par
The motivation of the present contribution is to introduce a method that allows the determination of the relative orientation of groups of atoms and reconstruct the grain from only this information. The method is able to generate a space-resolved three dimensional grain decomposition of atom-position data sets for FCC materials, such as those generated by MD-simulations. The first step of the method calculates an orientation for each atom by taking into account the positions of the nearest-neighbor atoms. In a second step, the method generates grain entities from atoms with similar orientations by collecting them via nearest-neighbor paths. The method is capable of calculating grain specific properties, namely volume, center of mass, average orientation and orientation spread. The method additionally provides a mapping method in order to track grains during a time-row data set. The method is introduced in order to be able to track the grain growth evolution of individual grains during MD-simulations, additionally enabling a grain-resolved visualization of polycrystalline FCC data sets. Whereas the algorithms here presented were independently developed, they are admittedly similar to the method introduced by Panzarino et al. \cite{Panzarino2015,Panzarino2014}. However, we emphasize that in contrast to these previous contributions, we developed our algorithms for high-performance post-processing and that we offer the code to the community as a totally open-source project \cite{gplv3}.
\section{Methods}

To identify the grains from a data set containing atomic positions, it is necessary to determine first the orientation of the atoms as a per-atom attribute from the local neighborhood. Once an atom (referred to as \emph{central atom}) is selected arbitrarily, the calculation of the orientation proceeds in the following way:

\begin{enumerate}
\item{Identify all the nearest neighbors and calculate the relative positions of the neighbors to the corresponding central atom.}
\item{Determine the affine $3 \times 3$ transformation matrix $M$ from the reference (non-rotated case) to the current crystal-configuration.}
\item{Determine the rotation that best fits $M$ as a least square solution.}
\end{enumerate}

All of these steps have an inherent mathematical and/or programmatic complexity. In the present section, we will discuss the algorithms that were utilized to solve each of these problems. In the appendix, these solutions are presented in the form of pseudo-code whereas the program is offered as an open-source project \cite{Hoffrogge01}.

\subsection{Orientation determination}
\label{OriDet}

The calculation begins with the identification of the nearest-neighbors. Once the neighbors of an atom have been identified, displacement vectors are calculated with the current atom as origin. These vectors are then normalized so that only unit vectors are used for the orientation calculation. Since in the FCC crystal structure each neighbor has an antipodal counterpart, the list of nearest neighbor vectors is reduced by finding vectors with a near $180^{\circ}$-relationship. The resulting mean direction $\vec{v_r}$ is calculated by vector subtraction of any $\vec{v}$ and its antipodal partner $\vec{v_a}$:

\begin{equation}
\label{Eq1}
\vec{v_r} = \dfrac{1}{2}(\vec{v} - \vec{v_a})
\end{equation}

In the FCC lattice, there are three pairs of neighbor vectors, each lying on different $\left\{100\right\}$ planes. Since each vector ought to have only one defined perpendicular partner in the FCC-case, we try to find the directions of this planes by identifying vector pairs with a near $90^{\circ}$-relationship. Now, at least six different vectors must be available in order to find all three pairs. Afterwards, the $\langle 100 \rangle$ directions $\vec{v_{100}}$ are calculated as the cross product of both pairs vectors $\vec{v_1}$ and $\vec{v_2}$:

\begin{equation}
\label{Eq2}
\vec{v_{100}} = \vec{v_1} \times \vec{v_2}
\end{equation}
The transformation matrix $M$ can now easily be obtained as:
\begin{equation}
\label{Eq3}
M=\left[\begin{matrix}
v_{x_{100}}&v_{y_{100}}&v_{z_{100}} \\
v_{x_{010}}&v_{y_{010}}&v_{z_{010}} \\
v_{x_{001}}&v_{y_{001}}&v_{z_{001}} \\
\end{matrix}\right]
\end{equation}

The assignment of the vectors to a certain row is of arbitrary nature, which can cause a negative determinant. It is possible to avoid this issue by calculating the determinant and checking whether it is negative or not. In the case that it results negative, the algorithm inserts $-\vec{v_{100}}$ in the first row instead of $\vec{v_{100}}$. The matrix $M$ is not necessarily a pure rotation matrix because the atoms can assume non perfect lattice positions due to elastic distortions or thermal vibrations. For this reason, we have to estimate the rotation that best fits the affine transformation defined by matrix $M$. This problem is well known in aeronautics and thus, several solutions have been already provided in the past. For instance, Horn \cite{Horn} introduced a direct method that estimates an orientation from a set of corresponding vectors in two coordinate systems for aeronautical purposes. A similar method that requires computing the eigenvalues of a $4 \times 4$ matrix was presented by Bar-Itzhack \cite{BarIt}. This method is based on the $q$-method proposed by Keat \cite{Keat01}. We opted to use this method in the present contribution as operations with quaternions are computationally less expensive. The $q$-method utilizes a cost function $c$ to be minimized:

\begin{equation}
\label{Eq4}
c = \frac{1}{2}\sum_{i=1}^{k}a_i |\vec{b_i} -D(q) \vec{r_i}|^2
\end{equation}

where $\vec{r_i}$ denotes a unit vector in the reference/sample coordinate system and $\vec{b_i}$ the same vector in the body/crystal coordinate system. In turn, $q$ is the sought rotation quaternion, which defines the rotation matrix $D$ whereas $a_i$ is an optional weight to each of the vector pairs. The cost function contains the square lengths of the error vectors, i.e. the minimization is applied with the Euclidean norm. In \cite{Keat01}, it was shown that the solution of this problem is reduced to the computation of the eigenvalues of a symmetrical $4 \times 4$ matrix, which is defined by the set of vectors  $\vec{r_i}$ and $\vec{b_i}$. To find the best-fitting rotation quaternion to the transformation matrix, the method introduced in \cite{BarIt} was utilized. In this contribution, it was shown that by utilizing the three base vectors $\left[100\right]$, $\left[010\right]$ and $\left[001\right]$ of the reference coordinate system, the problem can be formulated for non-orthogonal matrices. As a result, the new matrix $M_4$ can be computed from the values of the matrix $M$ from \cref{Eq3} as:

\begin{equation}
\label{Eq5}
M_4 = \frac{1}{3}\left[\begin{matrix}
r_{11} - r_{22} - r_{33} & r_{21} + r_{12} &  r_{31} + r_{13} &  r_{23} - r_{32} \\
r_{21} + r_{12} &	r_{22} - r_{11} - r_{33} & r_{32} + r_{23} & r_{31} - r_{13} \\
r_{31} + r_{13} &  r_{32} + r_{23} & r_{33} - r_{11} - r_{22}  & r_{12} - r_{21} \\
r_{23} - r_{32} &  r_{31} - r_{13} &  r_{12} - r_{21} & r_{11} + r_{22} + r_{33} \\
\end{matrix}\right]
\end{equation}

The quaternion $q$ representing the rotation that best fits $M$ is the eigenvector of the most positive eigenvalue of $M_4$. The determined quaternion $q$ is then assigned to the current atom during the calculation. In the case that the matrix $M$ cannot be defined owing to strong lattice distortions, the corresponding atom is marked as non-oriented.

\subsection{Grain detection}
After the assignation of the orientation to the atoms, the grains are identified in a second step. The identification method uses the orientation information generated in the previous step. The procedure begins from an initial atom as a seed and recursive assignations are applied to nearest-neighbor atoms. For this, the following steps are conducted:
\begin{enumerate}
\item{Check whether the nearest-neighbor atoms to a seed atom:}
\begin{enumerate}
\item{are not yet assigned to another grain-entity.}
\item{own an orientation close to the seed-atom according to a user defined threshold ($1^{\circ}$ as default). This is referred to as the \emph{local criterion}.}
\item{own an orientation close ($<3^{\circ}$ as default) to the mean orientation of the current grain if and only if enough atoms have been aggregated to this grain. This is referred to as the \emph{global criterion}.}
\end{enumerate}
\item{If these conditions are fulfilled, the neighbor atom is aggregated. The procedure is now repeated from the beginning with the aggregated atoms as seeds.}
\end{enumerate}
This procedure is exercised for every atom in the data set.  It can occur during an iteration that the number of aggregated atoms of a grain is smaller than a reasonable amount ($200$ atoms as default).  Whenever this happens, the corresponding grain information is removed and all collected atoms during this step are set to be unassigned again.

\begin{figure}[h!]
	\centering%
	\includegraphics[width=1.0\textwidth]{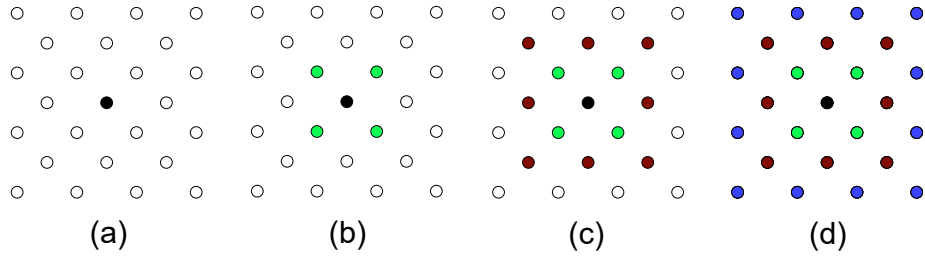}
	\caption{Sequence for the identification of the grains. (a) A seed atom is selected from the dataset; open circles represent atoms with the same orientation as the seed atom. (b) The first neighbors to the seed are identified by comparison of the per-atom orientations and collected into a grain entity. (c-d) In subsequent iterations, the already identified atoms (green in (c), brown in (d)) act as seeds for their next-neighbors. Atoms are continually collected until no equally oriented atoms can be found in an iteration step.}%
	\label{Fig1}%
\end{figure}

\begin{figure}[h!]
\centering%
\includegraphics[width=1.0\textwidth]{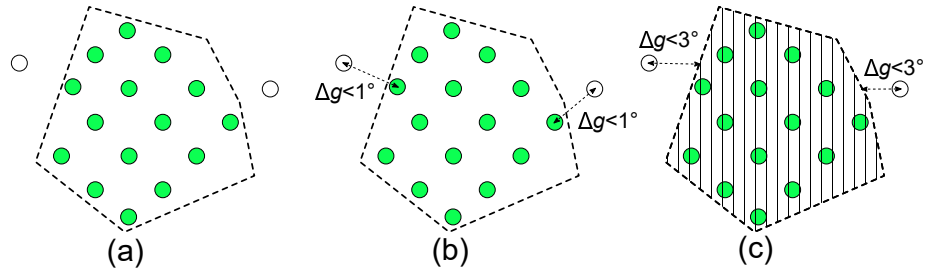}
\caption{A fully identified grain (a) consists of the atoms (solid circles) enclosed by the dashed line after few iterations. Nearest-neighbors to the already aggregated atoms are tested in subsequent iterations. In order to collect an atom to the current grain entity, the atom must have an orientation difference less than 1$\grad$ with respect to the seed atom (b) and additionally ensure that the orientation difference with respect to the mean orientation of the grain is less than 3$\grad$ (c).}
\label{Fig2}
\end{figure}

In \cref{Fig2}, the application of the misorientation criteria is exemplified. The \emph{local criterion} compares the orientation of two atoms in a nearest-neighbor relationship whereas the \emph{global criterion} assures that the orientation of an atom to be aggregated has a similar orientation than that of the current grain. The \emph{global criterion} allows for the detection of grains separated by low angle grain boundaries, which otherwise cannot be detected only by local comparisons due to the gradual change of orientation between grains at positions far away from dislocation centers.

\subsection{Orphan atom adoption}
\begin{figure}[!htb]
\centering%
\includegraphics[width=1.0\textwidth]{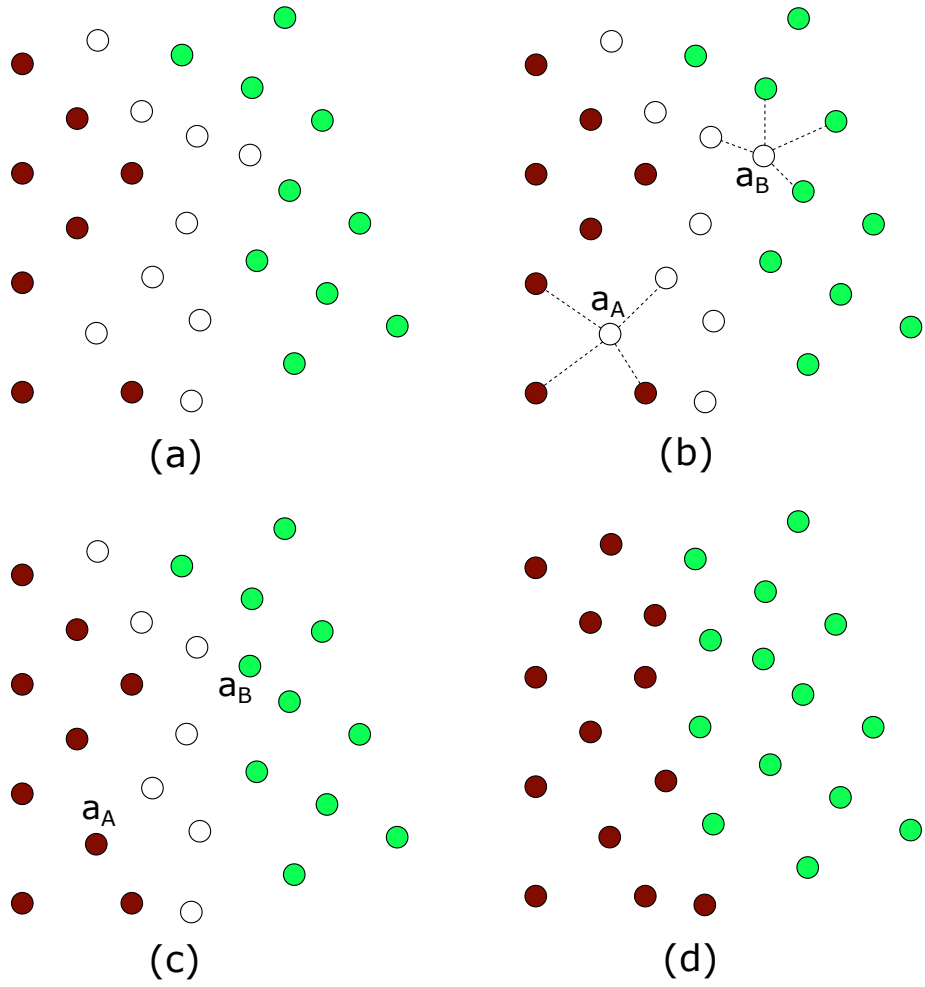}
\caption{Adoption of orphan atoms. The orphaned state is represented by open circles whereas differently colored circles signify atoms aggregated to different grains. (a) Initially, many atoms close to grain boundaries are not assigned to a specific grain. These atoms are adopted recursively depending on the frequency occurrences to neighboring grains (b). For example, in (c) atom $a_A$ is adopted by the grain on the left whereas atom $a_B$ is adopted by the grain on the right. After a few iterations the number of orphan atoms is reduced to a minimum (d).}
\label{Fig3}
\end{figure}

In some cases owing to strong thermal vibrations or proximity to lattice defects, the determination of the orientation can fail or deliver orientations different than those found in an immediate neighborhood. In both cases, such atoms will not be assigned to a grain entity during the grain identification step. These atoms, deemed in the following as \emph{orphan atoms}, can cause a substantial underestimation of the volume of the grains. To solve this problem, we utilized a method inspired in noise correction methods used in EBSD mapping algorithms \cite{Humphreys01, Humphreys02}. The approach consists of the following steps:
\begin{enumerate}[label=(\alph*)]
\item{For each orphan atom, the grains to which the nearest neighbors belong are identified and counted.}
\item{From the list of grains determined in the previous step, the one with most occurrences is selected.}
\item{If the maximum number of occurrences is higher than a user defined value (the default is three), the orphan atom is assigned to the grain selected in step (b).}
\item{The procedure is repeated for all remaining orphan atoms.}
\end{enumerate}
The minimum number of grain occurrences (three) was introduced in order to suppress random grain assignations. This procedure is schematically shown in \cref{Fig3}. Note that initially (\cref{Fig3}b) only orphan atoms $a_A$ and $a_B$ are adopted because only they had enough nearest neighbors assigned to a common grain. This procedure is applied iteratively so that the number of orphan atoms is reduced to a minimum. 

\hl{The adoption procedure is expected to introduce an error to the shape of the grain boundaries and triple lines as the assignation of the atoms is not anymore perfectly deterministic. Nevertheless, this error is estimated to be negligible as its possible magnitude is in the range of the lattice parameter whereas the dimension of triple lines and grain boundaries are several times that scale and is comparable to the grain size. For an exact quantification of the error, the identification of the structural elements of the microstructure is necessary. This feature is not yet implemented in the program.}

\subsection{Grain-tracking over time}
In order to determine the evolution of grain properties over time, it is necessary to track the grains from snapshots generated at different times. To do this, we assume that geometrical and orientation attributes of the grains show little change between subsequent time states. This enables finding correspondence of grains by comparing calculated properties. We utilized two criteria to identify possible corresponding candidates, namely the center-of-mass (COM) and the misorientation between grain candidates.
To begin with, the COM of the grains is calculated for the data sets generated at times $t$ and $t+\Delta t$. An arbitrary grain $G_1$ is selected from the second data set $(t+\Delta t)$. From the first one ($t$), the grains that fulfill $d \leq d_{max}$, where $d$ is the distance to grain $G_1$ and $d_{max}$ is a user defined threshold value, are chosen. This list is further refined by rejecting grains with a misorientation larger than a user-defined value ($\Delta g>5\grad$ as default). If more than one grain exist that fulfill both criteria, the grain with the shortest distance $d$ is chosen.

The methods described in the previous sections are provided as pseudo-code in the appendix and implemented as a command-line program called GraDe-A (Grain Detection Algorithm). The program was written in the object-oriented C++ programming language. Furthermore, the software uses shared-memory parallelization by means of the OpenMP-API \cite{dagum1998openmp}, which enables analyzing several input files in parallel. The code has been written accounting for computation time considerations. Thereby, the utilization of computationally slow operations such as square root or cosine calculations has been avoided. Additionally, the computation domain has been split into cells in order to accelerate nearest-neighbor-search algorithms. All orientation data is internally stored as unit quaternions with memory consumption for only four floating-point numbers. Therefore, all misorientation calculations are implemented as simple quaternion operations, which are superior to rotation matrix operations with regard to computational effort. Unique cubic orientation quaternions (see \ref{alg:CubUniqueOrients}) \cite{Cho2005} are used in order to be able to calculate the mean-orientation and the orientation spread \cite{Cho2005} for each grain. Additionally, this enables to use a fast disorientation calculation for the grain identification algorithm. The code is also offered as open-source software \cite{Hoffrogge01}.

\subsection{Molecular dynamics simulations of grain growth}
Molecular dynamics simulations were employed to test the developed software. For this purpose, nanocrystalline grain growth was simulated. The large-scale atomic/molecular massively parallel simulator (LAMMPS) code \cite{Plimpton01} was utilized. The atomic interactions were described by the EAM potential for Al developed by Mishin et al. \cite{Mishin02}. The algorithms presented in previous sections were used to determine the orientation of grains in an Al polycrystal and reconstruct them in time-resolved molecular dynamics data sets \cite{Hoffrogge01}. The simulation block used in these MD simulations was composed of one hundred Voronoi grains with different orientations (\cref{Fig4}) that were defined so that a weak texture was predominant in the microstructure. The orientations were assigned randomly to the grains. The simulation box had a side length of $43.27$ nm in all directions and contained $4834137$ atoms for an initial average grain size of approximately $11.57$ nm. Periodic boundary conditions were used on all the surfaces of the simulation box. Before the MD simulation was performed, the energy of the system was minimized by the conjugate-gradient algorithm. Subsequently, damped dynamics were applied to fully relax the grain boundaries. The isothermal-isobaric (NPT) ensemble was used for the time integration with a time step of $0.2$ fs. The simulation was performed at $500$ $\grad$C. The total simulation time corresponded to $4$ ns. In order to warrant the accuracy of the measured properties of the simulated systems, snapshots were regularly saved in intervals of $0.01$ ns. The obtained configurations were subsequently quenched computationally in an NPT-ensemble for a time of $0.1$ ns to a target temperature of $1$ K and additionally relaxed by the conjugate gradient algorithm after setting the kinetic energy to zero. For the identification of the grains, the default thresholds were consistently used for all of the data analyzed. An amount of $200$ atoms was set as a minimum for any grain detection.

The MD-simulations were performed on the J\"ulich Supercomputing Cluster and the post-processing of the data on the RWTH-Aachen Computer Cluster.
\section{Results}

\subsection{Validation and computational performance}

The algorithm was benchmarked on a Bull MPI-S machine possessing Intel Xeon X5675 processors. For these tests, we used $24$ files of approximately $240$ MB containing $\sim$4.8 million atoms. The utilized parallelization scheme allowed reducing the computation time from 1.02 hours for two threads to only 12 minutes for 12 threads (\cref{Tab1}). Irregular scaling with the number of threads was observed. This behavior was caused by an asymmetrical distribution of tasks per thread seen in situations where the number of tasks is not a multiple of the number of threads e.g. 10 threads. This problem is less significant with an increasing number of tasks. For instance, in the present contribution a total of 400 files for a total data volume of 140 GB were evaluated in only approximately 4 hours. 

\begin{table}[htb!]
	\centering
	\caption{Comparison of the total computation time $t_{comp}$ necessary for processing 24 output-files by the Grade-A tool. A different number of threads (\#Th) were used for benchmarking in a linux-environment with 12 CPU-cores. Each file contained about 4.8 million atoms and had a size of approximately 240 MB. The relative overhead was calculated using the computation with two threads as reference.}
	\begin{tabular}[h]{llll}
	\#Th& $t_{comp} (h)$	& $t_{comp}\times$\#Th (h)	& Overhead $(\%)$\\
	\hline
	2&	1.02&	2.03&	0.00\\
	4&	0.53&	2.12&	3.78\\
	6&	0.37&	2.23&	9.469\\
	8&	0.29&	2.28&	12.17\\
	10&	0.28&	2.77&	36.10\\
	12&	0.20&	2.43&	19.29\\
	\end{tabular}
	\label{Tab1}
\end{table}

\begin{figure}[h!]
\centering
\subfloat[]{%
  \includegraphics[width=0.75\textwidth]{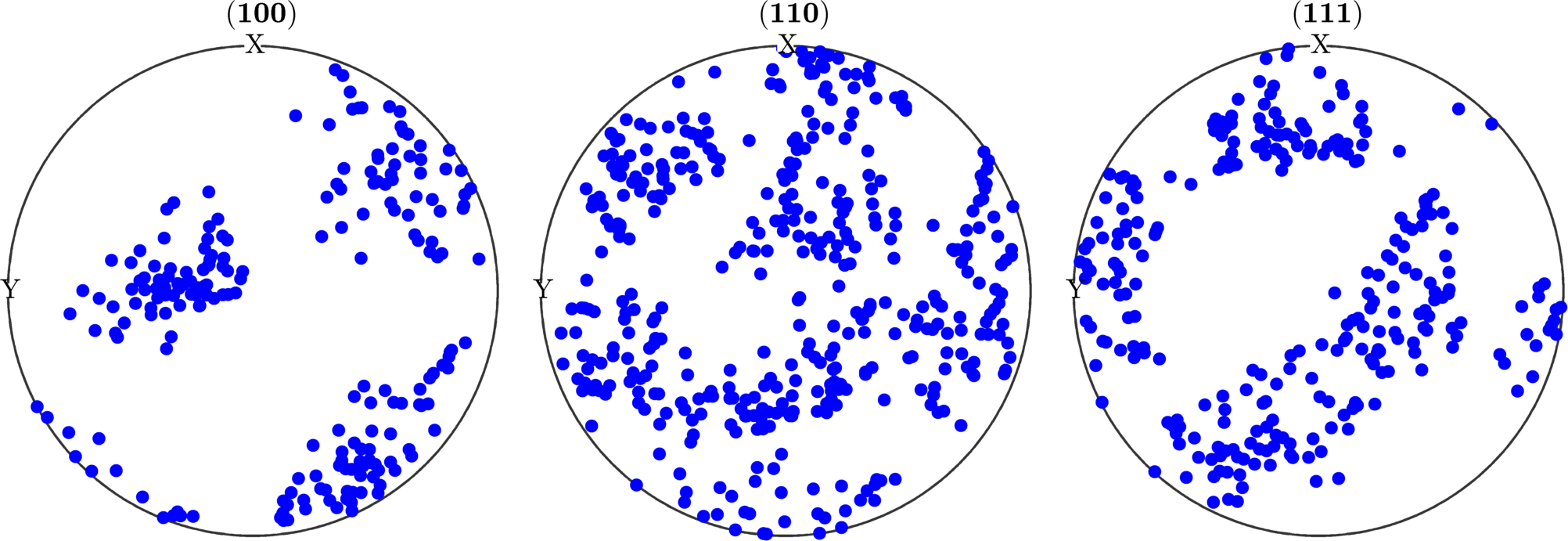}%
  }\par
\subfloat[]{%
  \includegraphics[width=0.75\textwidth]{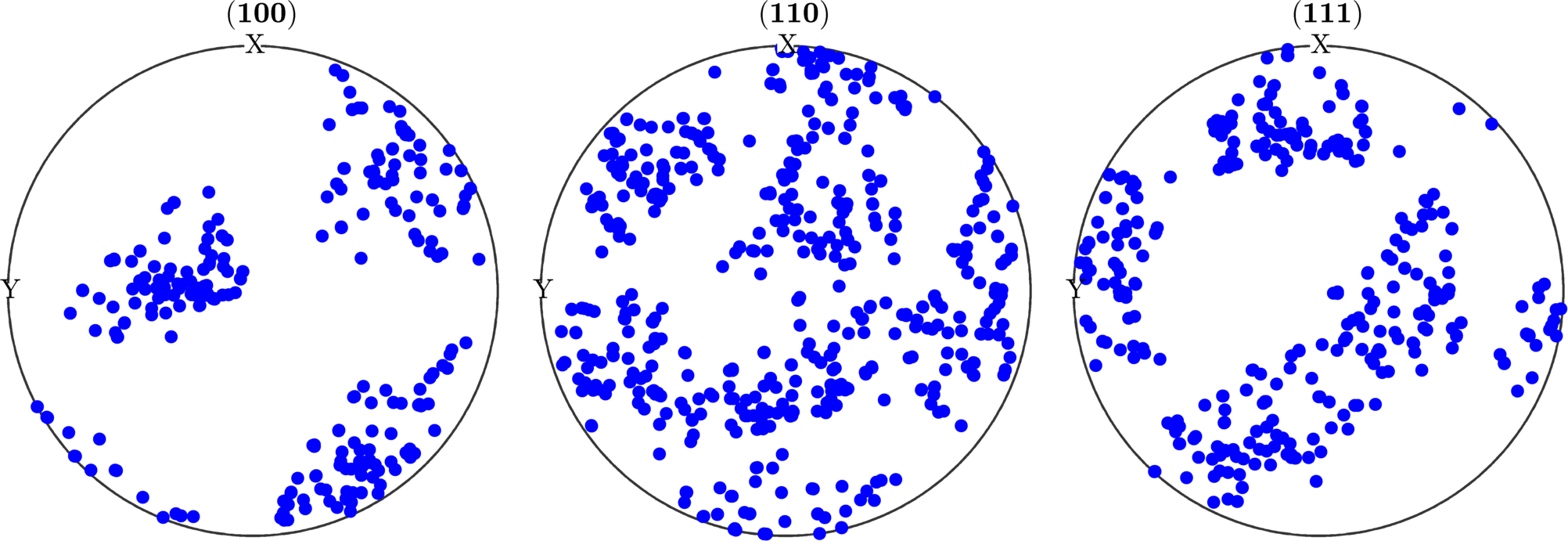}%
  }\par        
\subfloat[]{%
  \includegraphics[width=0.40\textwidth]{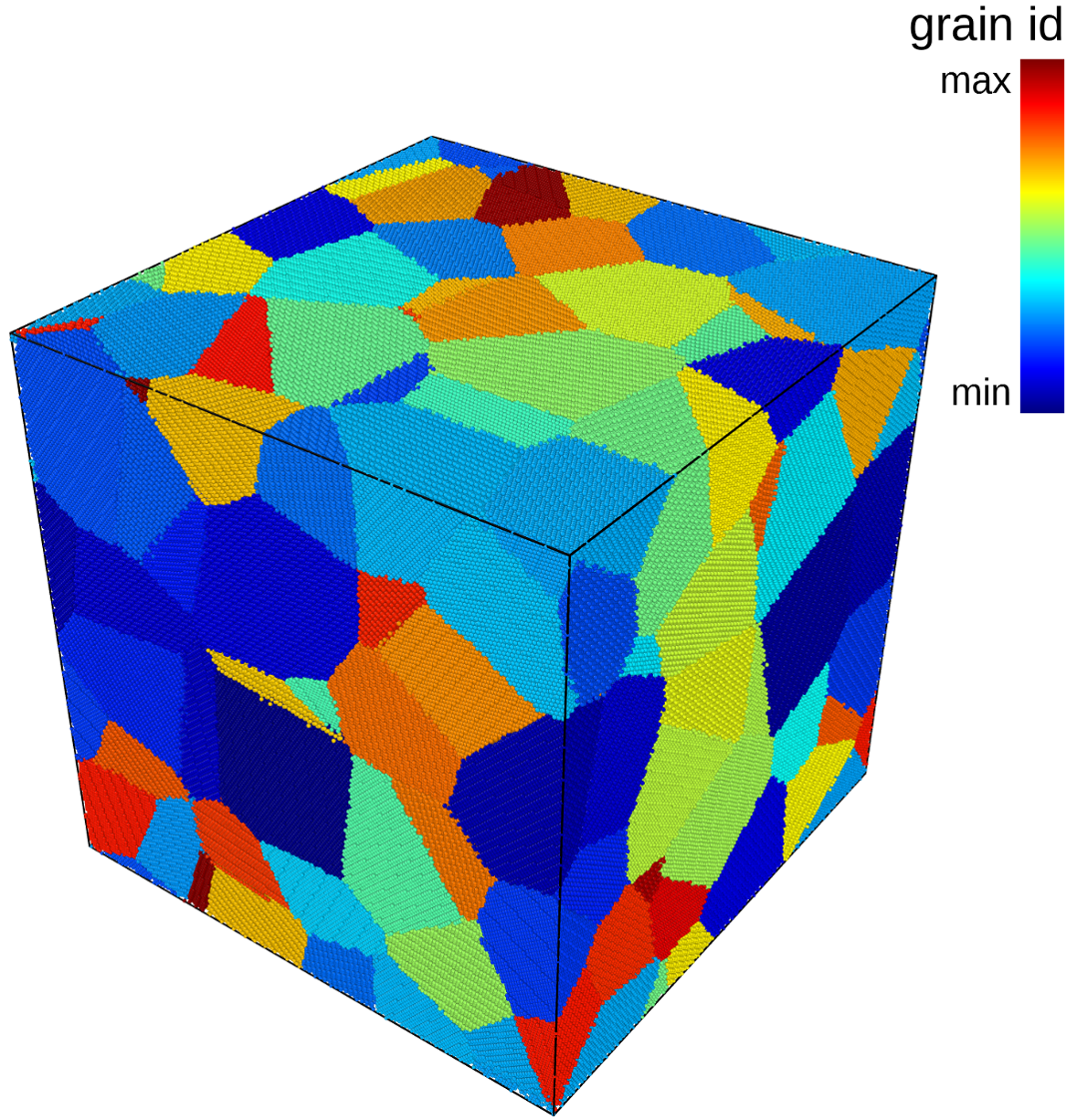}%
  }
\caption{(a) The input orientation are shown in the pole figures and (b) compared to the determined orientations by the GraDe-A tool. An excellent agreement can be observed. The reconstruction of the grains is seen in (c), where the grains are colored according to an integer number that serves as an id for the grains. This number was assigned so that the smallest number corresponds to grains with the largest volume and vice versa.}
\label{Fig4}
\end{figure}

To estimate the accuracy of the algorithm, the determined orientations were compared to the ones used as input. It is stressed that the algorithm does not require initial orientations as input. The figures plotted in \cref{Fig4} illustrate clearly that the algorithm is capable to identify correctly the orientations of the grains. \cref{Fig4}a and \cref{Fig4}b correspond respectively to the seeded orientations and the orientations calculated from the MD data set. \cref{Fig4}c shows the microstructure colored after the identification number of the grains, which is a number determined by the sorted volume of the grains. This is the reason, why large grains appear blue and small red. The identification number can change from one snapshot/time-step to another due simply to the growth and shrinkage of the grains. However, as explained in the previous section the algorithm is capable of tracking correctly the grains as the COM and orientation are the criteria utilized for the tracking, independently of the identification number assigned at a time step.

\begin{figure}[htb!]

\begin{minipage}{.5\linewidth}
\centering
\subfloat[]{\label{main:a}\includegraphics[width=1.0\textwidth]{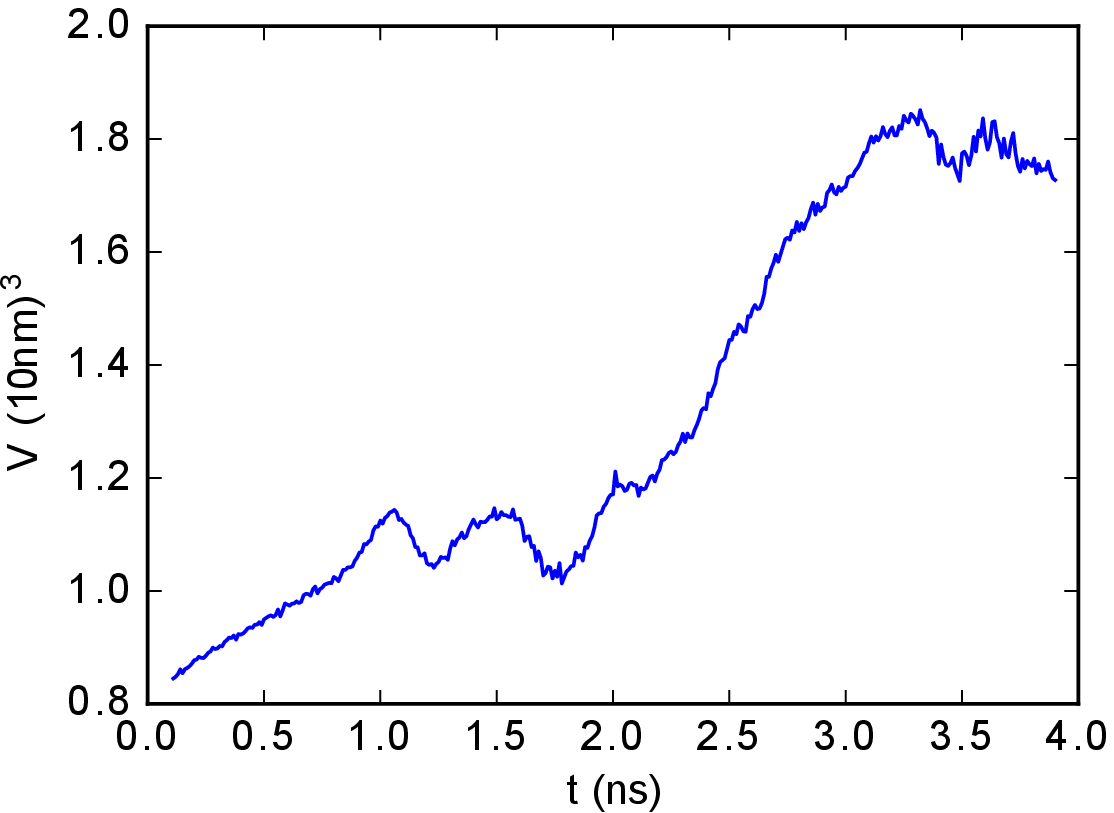}}
\end{minipage}%
\begin{minipage}{.5\linewidth}
\centering
\subfloat[]{\label{main:b}\includegraphics[width=1.0\textwidth]{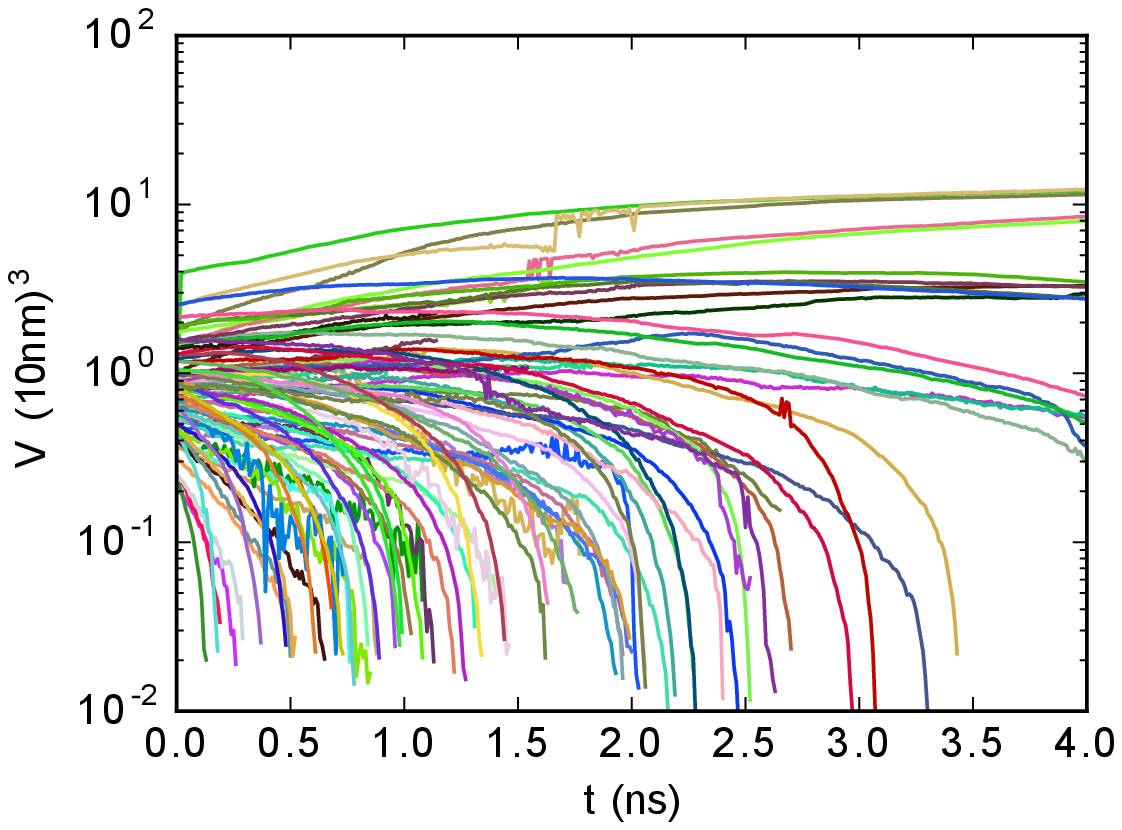}}
\end{minipage}\par\medskip
\centering
\subfloat[]{\label{main:c}\includegraphics[width=0.8\textwidth]{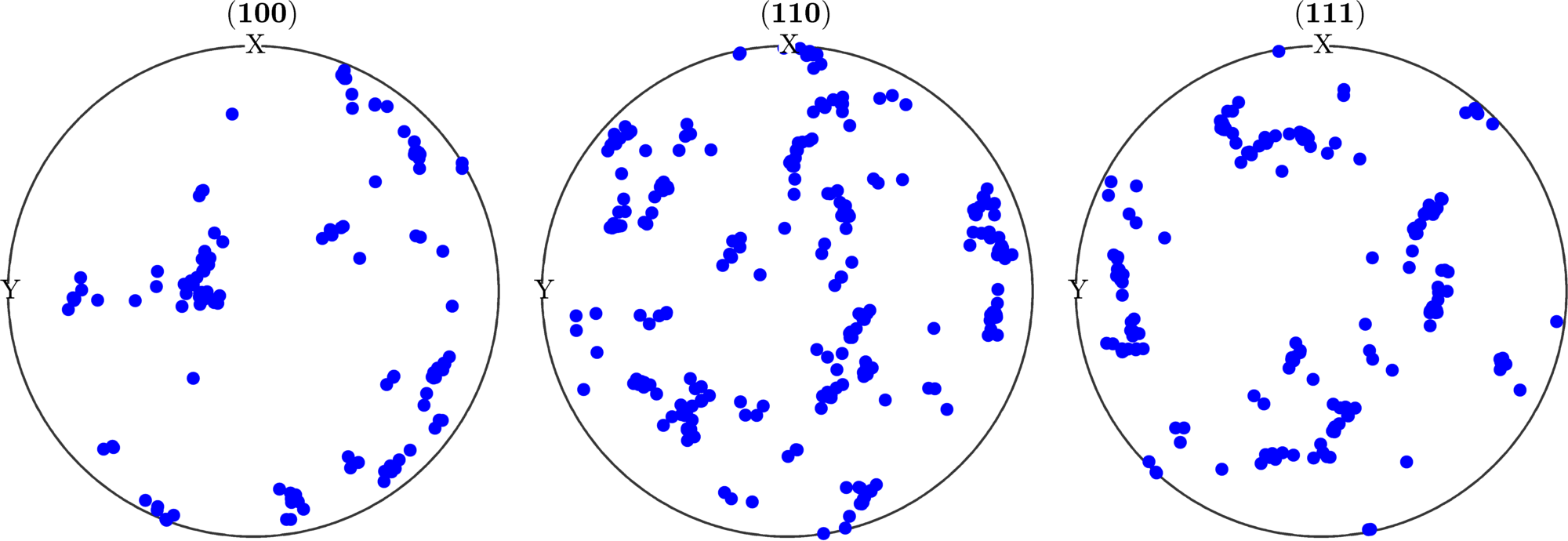}}

\caption{(a) Global kinetics, (b) individual kinetics and final texture (c) after 4 ns of computational annealing at 500 $\grad$C. \hl{Fluctuations observed in the development of the kinetics are caused by the reconstruction algorithm and thermal fluctuations as grains with approximately 200 atoms can suddenly gain or lose atoms and thus be discriminated as either valid or invalid grains. It is noted that the moving average is shown in (a); as the amount of grains is small, the fluctuations appear more pronounced.}}
\label{Fig5}
\end{figure}

Clearly, the advantage of the grain reconstruction is that it makes possible the tracking of grain properties such as volume, center of mass, average orientation, orientation spread and orientation change. Additionally, the algorithm permits the identification of sudden events that influence decisively the development of the microstructure. For instance, \cref{Fig5}a illustrates the global kinetics obtained by detecting the remaining grains at each snapshot of the simulation. It is noted that since grains are individually identified, other algorithms for grain size measurements, such as fitting ellipsoids to the grains \cite{Groeber}, can be easily implemented.

\subsection{Grain growth kinetics and rotation}

\begin{figure}[h!]

\begin{minipage}{.5\linewidth}
\centering
\subfloat[]{\label{main:a}\includegraphics[width=1.0\textwidth]{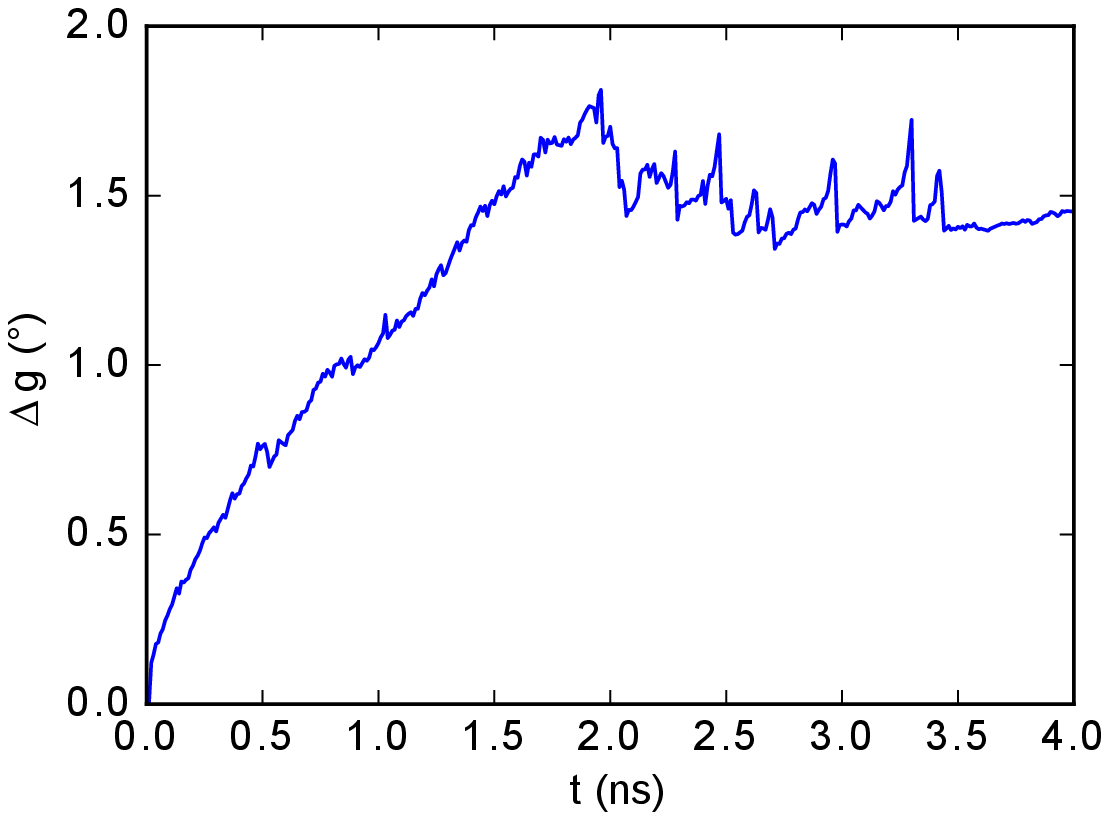}}
\end{minipage}%
\begin{minipage}{.5\linewidth}
\centering
\subfloat[]{\label{main:b}\includegraphics[width=1.0\textwidth]{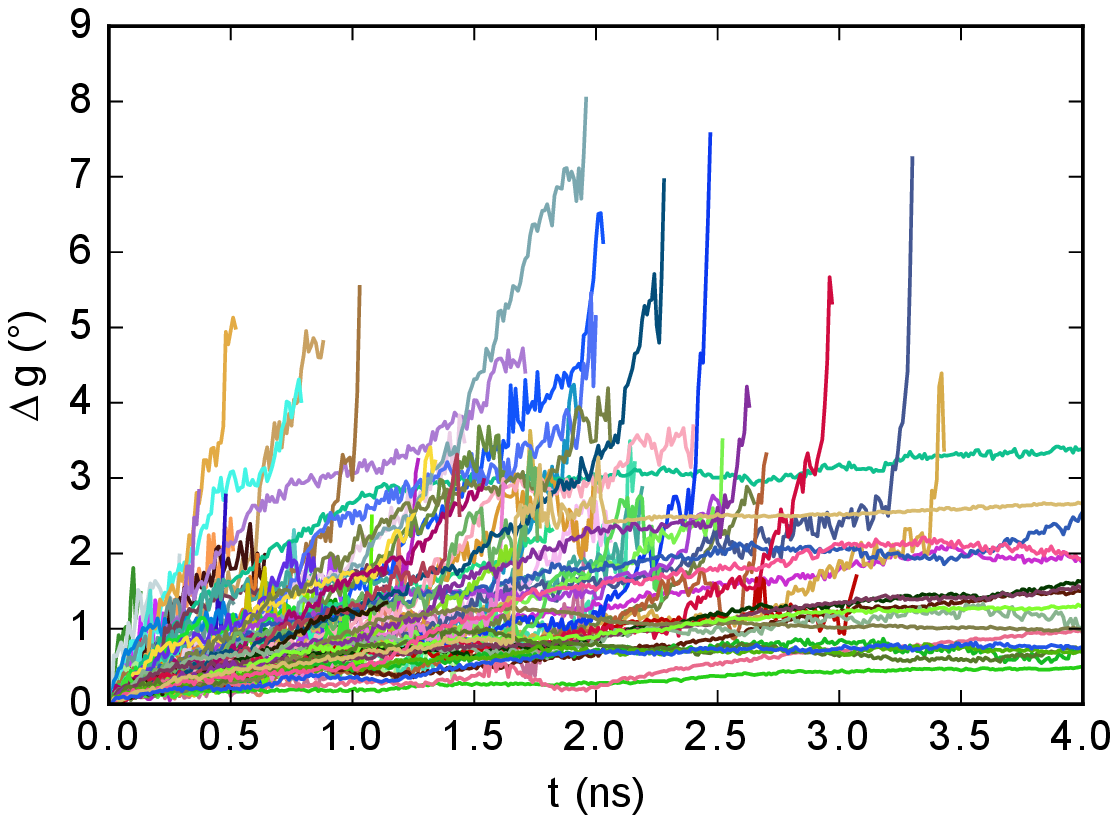}}
\end{minipage}\par\medskip
\begin{minipage}{.5\linewidth}
\centering
\subfloat[]{\label{main:c}\includegraphics[width=1.0\textwidth]{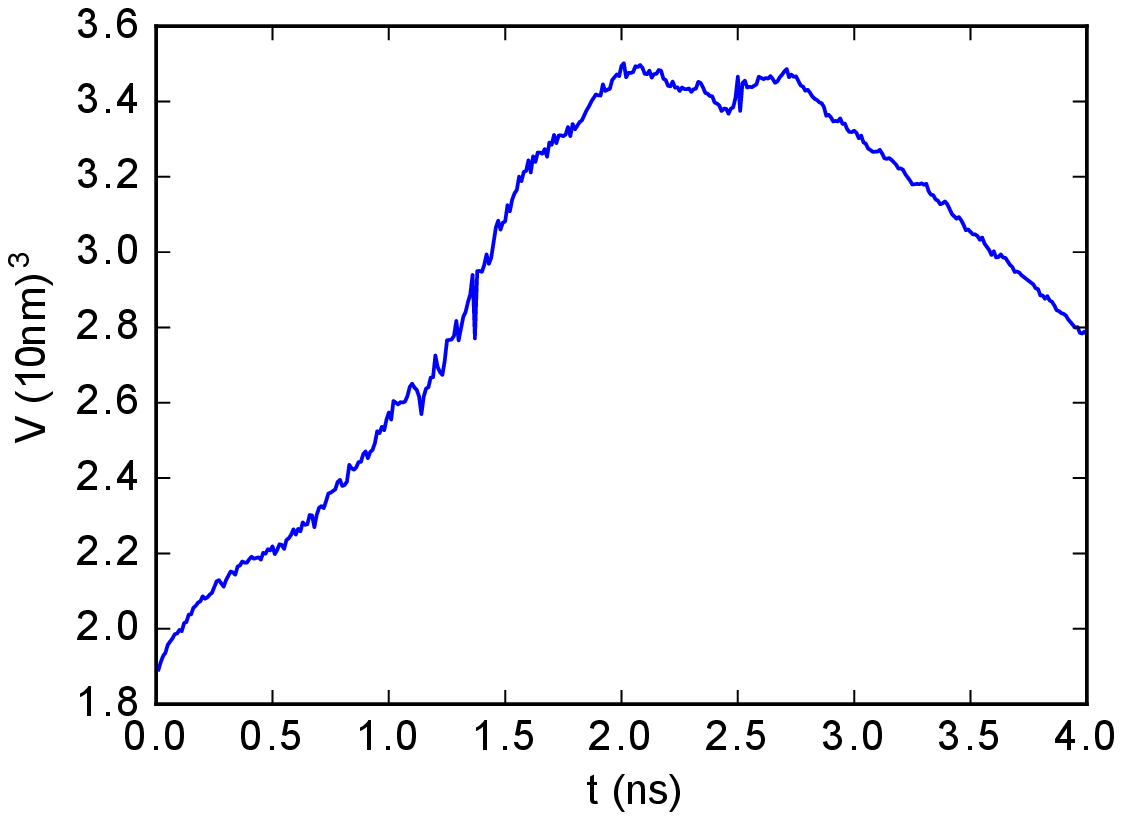}}
\end{minipage}
\begin{minipage}{.5\linewidth}
\centering
\subfloat[]{\label{main:d}\includegraphics[width=1.0\textwidth]{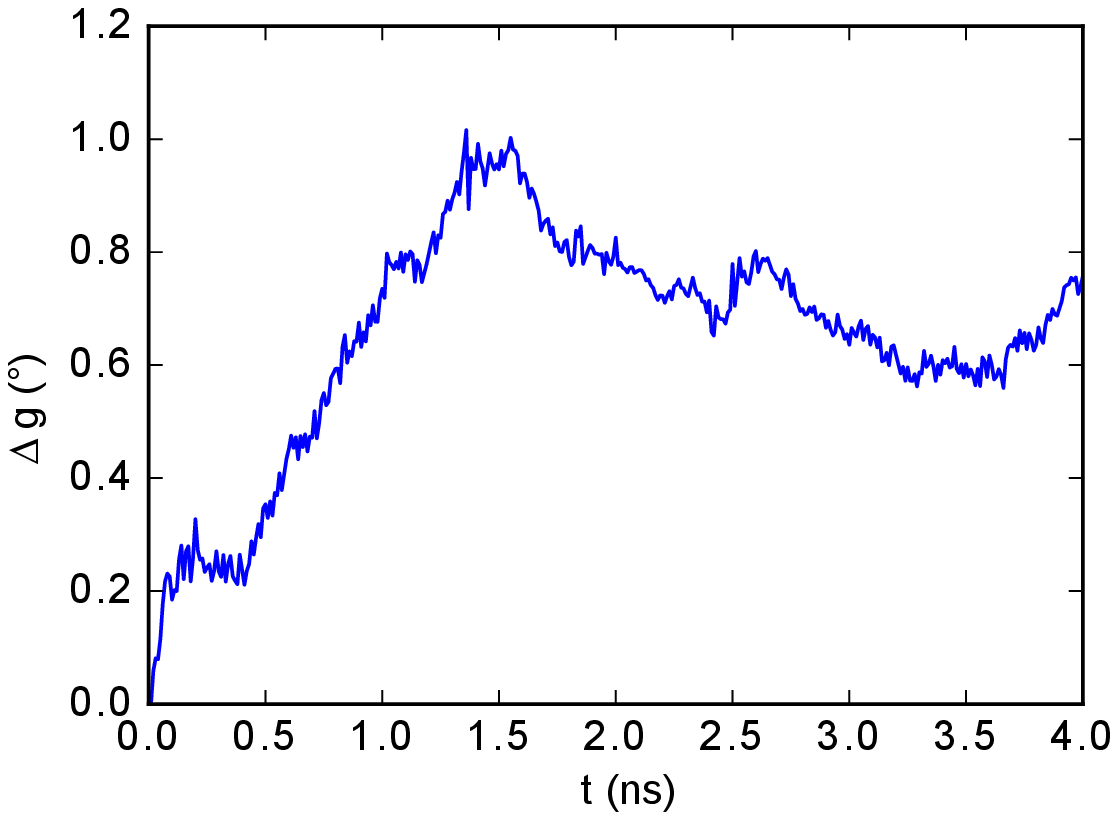}}
\end{minipage}

\caption{(a) Average grain rotation and (b) individual rotation of the grains. Grain rotation seems to increase with decreasing grain size of the grains. However, the average orientation change of the grains remains below 3.5$\grad$. Significant events for microstructure evolution can also be identified since properties such as kinetics (c) and orientation change (d) of the grains are individually resolved. In (c) and (d), the properties plotted correspond to the grain shown in \cref{Fig7}}.
\label{Fig6}
\end{figure}

\hl{From initial 100 grains in the microstructure, the number of grains decreased to only 40 after $4$ ns.} \Cref{Fig5}a \hl{reveals the average of the grain volume with time whereas} \cref{Fig5}b \hl{shows the individual kinetics of the detected grains.} Note that the ordinate is in logarithm scale so that the collapse of the grains is better visualized. For comparison, the orientations of the remaining grains after the finalization of the simulation are plotted in the pole figures shown in \cref{Fig5}c; a slight sharpening of the initially used weak texture is evident as expected for the texture evolution after grain growth \cite{Diss}. \hl{Remarkably, the evolution of the grains seems to agree qualitatively very well with simulations performed by mesoscopic simulation models} \cite{Zoellner2012}\hl{, where the kinetics of individual grains were also compared. In addition, the individual volume evolution of the grains is in accordance to theoretical expectations} \cite{Glicksman2007,Glicksman2005,Glicksman2005a,Glicksman2007a,Glicksman2009,Rios2006b,Rios2006c}\hl{. However, differences become evident, when the evolution in time of the grain volume is analyzed in detail as discussed below.} Since the algorithm is capable of determining a per-atom orientation and identify the grains, it is also possible to determine any change in orientation that might be induced during grain growth. This is a topic of interest because Cahn and Taylor \cite{Cahn01} predicted that rotation of grains can be caused by the motion of grain boundaries. Several simulation studies \cite{Trautt01, Trautt02, Trautt03, Barr01, Upmanyu01} and some experimental ones \cite{Harris1998,Legros01,Mompiou01} have confirmed this effect for stressed nanocrystalline polycrystals. However, other experimental studies \cite{Mompiou01} have not observed this concomitant grain migration and rotation in systems closer to the simulated ones, where rotation is thought to be more prominent. In fact, in previous simulation studies \cite{Barr01, Brandenburg01}, we showed that rotation is strongly dependent on the character of the grain boundary. Furthermore, most of the computational studies have been performed in bi- or tricrystals \cite{Trautt03} and with CSL grain boundaries as a necessity to fulfill periodicity on all surfaces of the simulation setups. However, grain boundaries in real microstructures tend to be of a mixed and general character and can deviate substantially from CSL or $\Sigma$ grain boundaries. For this reason, it is unclear whether grain rotation will be observed as a generalized mechanisms of microstructure evolution.
To investigate if grain rotation is observed in nanocrystalline polycrystals subjected to grain growth, we tracked the orientation change of the grains. In \cref{Fig6}a the average absolute change of grain orientation as a function of time is shown. The maximum orientation change was less than $3.5\grad$; this evinced only a marginal orientation change that is not in accordance with Cahn and Taylor theoretical expectations \cite{Cahn01, Barr01, Brandenburg01, Trautt02}. The reason for this may be related to the structure of the grain boundaries as discussed in \cite{Barr01} but a more detailed investigation of the mechanisms of grain boundary migration is necessary to affirm this. As for the kinetics, the change of orientation can also be resolved for individual grains. This is illustrated in \cref{Fig6}b, where the rotation of all the grains in the simulations is plotted as a function of time. Interestingly, rotation is most prominent for grains that are close to collapse as evinced by those grains, whose orientation change increases in a very short time. An important difference with respect to the simulations by Haslam et al. \cite{Haslam01} is that grain coalescence was not substantially observed. The reason for this is that Haslam simulated quasi-2D grain growth. In addition, all the grain boundaries in their microstructure were tilt boundaries, in which case, grain rotation is inevitable as discussed in \cite{Barr01}. In the present simulations most of the grain boundaries possess mixed grain boundaries as a result of the dimensionality of the microstructure, which results in a microstructure closer to real polycrystals. The effect of a mixed component of the grain boundary on grain rotation has been studied previously \cite{BarralesMora2016,Barr01}. These studies indicated that mixed grain boundaries with common $<100>$ rotation axis do not rotate by the capillary motion of the boundary. The fact that in our simulations no substantial rotation was observed seems to support this previous conclusion.
\par
\begin{figure}[h!]

\begin{minipage}{.5\linewidth}
\centering
\subfloat[]{\label{main:a}\includegraphics[width=0.75\textwidth]{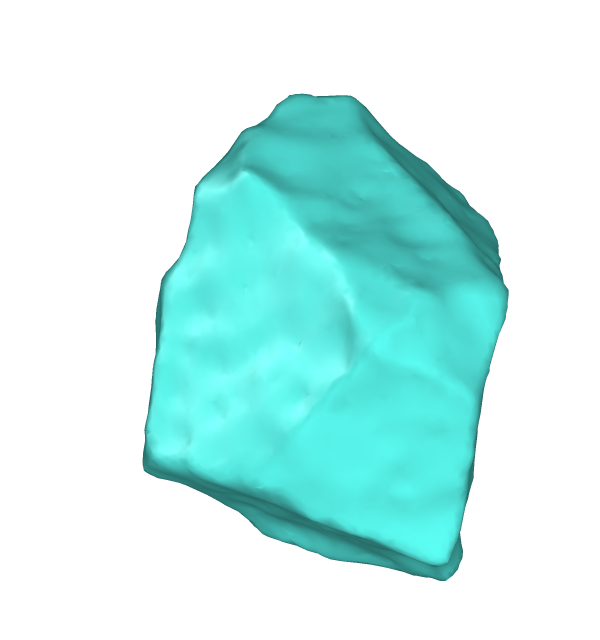}}
\end{minipage}%
\begin{minipage}{.5\linewidth}
\centering
\subfloat[]{\label{main:b}\includegraphics[width=0.75\textwidth]{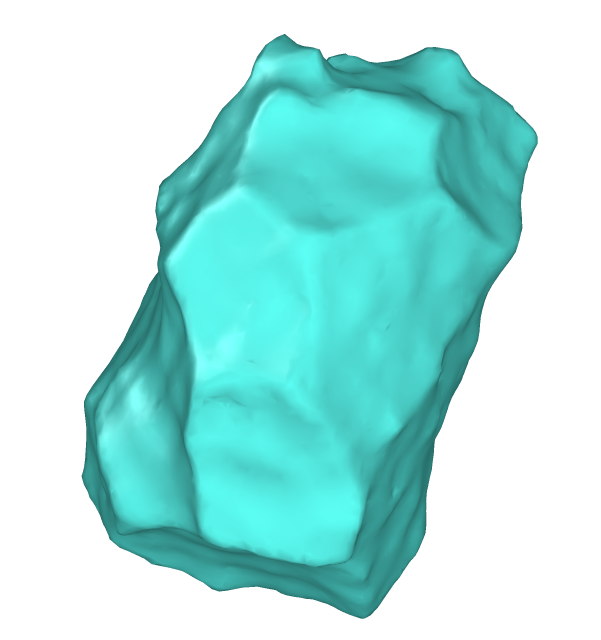}}
\end{minipage}
\begin{minipage}{.5\linewidth}
\centering
\subfloat[]{\label{main:c}\includegraphics[width=0.75\textwidth]{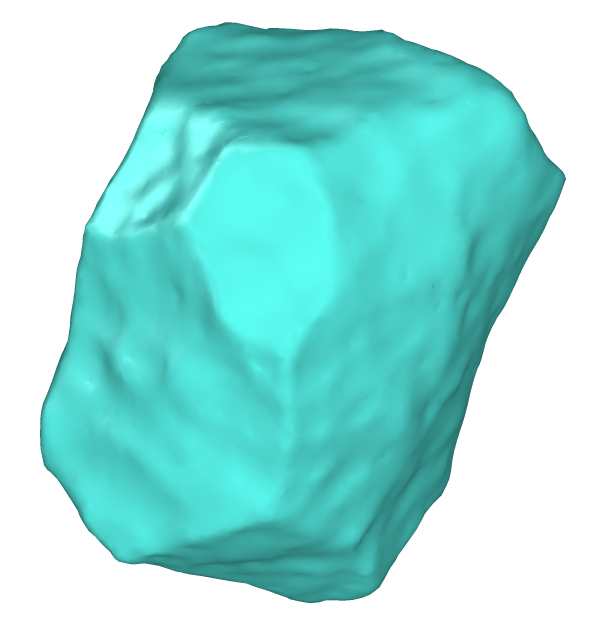}}
\end{minipage}
\begin{minipage}{.5\linewidth}
\centering
\subfloat[]{\label{main:d}\includegraphics[width=0.75\textwidth]{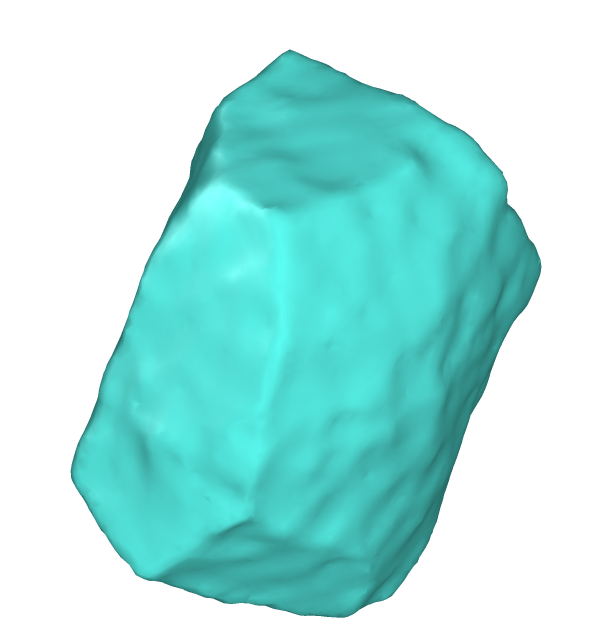}}
\end{minipage}

\caption{Development of an arbitrary grain as reconstructed by the GraDe-A tool. The surface was calculated by means of the visualization software OVITO \cite{0965-0393-18-1-015012}. This grain initially grew (b) until it reached a maximum number of faces after $1.33$ ns and a maximum size (c) after $2.66$ ns. Afterward, the grain shrank (d) until the simulation finalized after $4$ ns.}
\label{Fig7}
\end{figure}

Finally and to give an impression of the practicality of the algorithm, individual grain kinetics and orientation change for an arbitrary grain are shown in \cref{Fig6}c and \cref{Fig6}d. This representation makes possible identifying significant but rare events that are determinant for microstructure evolution \cite{Haslam01, Haslam02, Haslam03, Haslam04}.  In the case of this grain, the kinetics showed an initial fast growth until it reached a constant volume from approximately $2$ to $2.5$ ns. Afterward, the grain shrank until the simulation finished. The morphology of this grain can be seen in \cref{Fig7}. This grain was able to survive the whole simulation and grew successfully for most of its life reaching a maximum grain size after $2.66$ ns (\cref{Fig7}c). This growth is also substantiated by the number of faces that the grain gained. For instance, after $1.33$ ns (\cref{Fig7}b) the grain had significantly more faces than originally (\cref{Fig7}a) and evidently a concave shape that signifies positive growth rates. At the last simulated time (\cref{Fig7}d), the grain had already started shrinking showing a smaller size than at a previous time (\cref{Fig7}c) and also a convex shape. Note that the shape is not irregular and approximates excellently grains observed in polycrystals by serial-sectioning or tomography \cite{Rhines, McKenna, Moebus}. This is also an indication that the software can identify and reconstruct the grains adequately. 

\begin{figure}[h!]
\includegraphics[width=1.0\textwidth]{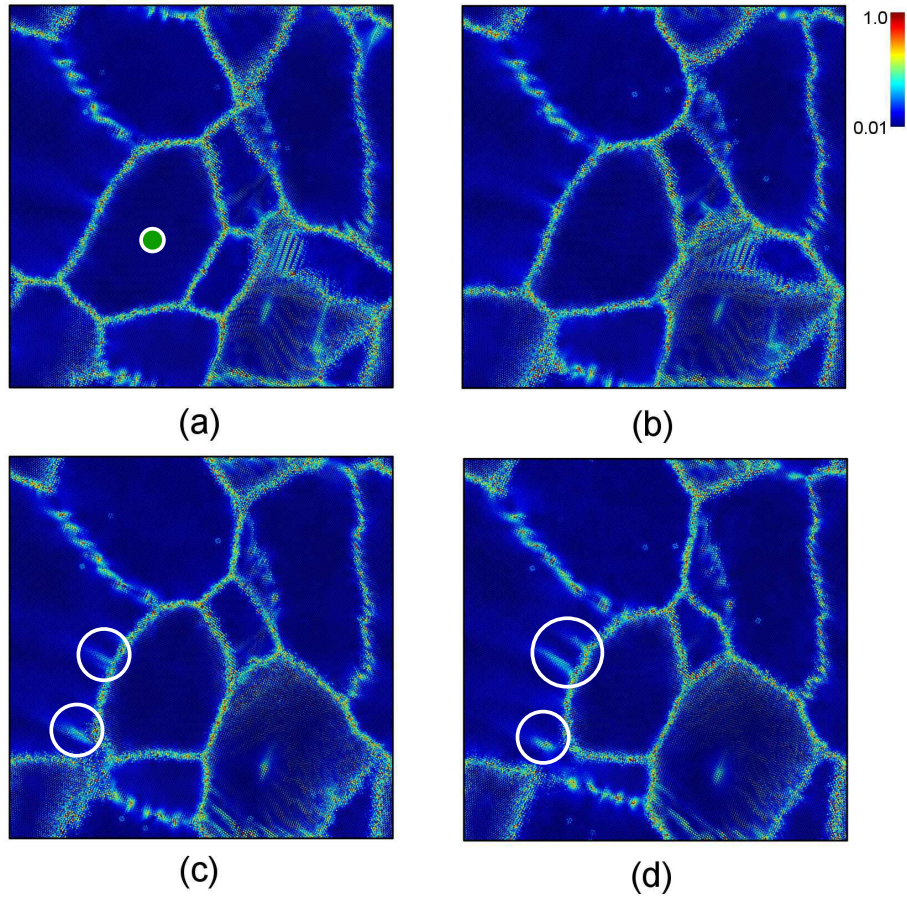}
\caption{(a) The grain with the solid circle in its interior corresponds to the same grain in \cref{Fig7} after $t=2.23$ $ns$. At this time, the growth rate is constant. Little change volume is observed even after $t=2.65$ $ns$ (b). One of the causes is apparently the dragging effect of grain boundary dislocations (white circles) (c) that must be emitted from the grain boundary before boundary migration can proceed at faster rates (d). In this figure, the grains were colored according to the von Mises stress in the grains. It is evident that no substantial changes of stress inside the grains is observed.}
\label{FigvonMisses}
\end{figure}

The stagnation of the growth of this grain is also evident at about $t \approx 2.0$ $ns$. It is impossible to separate the topology of the grain to the growth rate of the grain \cite{MacPherson2007}. However, it is also evident from the global kinetics (\cref{Fig5}) that grain growth did not occur ideally, which indicated a deviation from the theoretical expectations. \hl{Owing to the low number of grains in the microstructure, the observed tendencies and in particular the deviations may be caused by an insufficient sample size. However, in the simulations it was always possible to associate the retardation events to specific occurrences of grain boundary migration. For instance,} an inspection of the microstructure during grain growth revealed that also inherent features of the grain boundaries such as grain boundary dislocations can affect grain boundary motion and thus, grain growth. From the sequence shown in \cref{FigvonMisses}, it was possible to associate the stagnation of grain growth to the emission of dislocation from one of the grain boundaries in addition to the change of topology that occurred at a later time (\cref{FigvonMisses}c and \cref{FigvonMisses}d). For a discrimination of topological effects, the determination of the topology of the grains is necessary. This task is not trivial as not only the neighborhood of the grains but also their metrics must be determined \cite{MacPherson2007, Chang2012}. \hl{More intriguing is the effect of the initial microstructure, which was generated from a Voronoi tessellation. It is well known that the grain size distribution of Voronoi mosaics deviate strongly from the equilibrium distribution of normal or ideal grain growth. It might, of course, be argued that under experimental and the used simulation conditions no ideal grain growth can be expected. However, a transient from the initial to the equilibrium distribution can still be anticipated. Previous investigations} \cite{Barrales-Mora2012,Zoellner2016}\hl{ by mesoscopic simulations have substantiated that an initial non-equilibrium grain size distribution can considerably affect the evolution of grain growth especially in nanocrystalline materials. For instance, the increase of faces during the growth of the grain in} \cref{Fig7} \hl{is a feature that was already observed in simulations of early regimes of grain growth} \cite{Zoellner2016}\hl{. Nevertheless, from only one case of study, it is difficult to determine the extent of the effect of the initial microstructure. To study this and also the possible effects caused by the dimensionality of the sample, MD simulations of the same microstructure but with different initial grain size would be necessary. These simulations are at the moment being performed and we expect to report on them soon.}

The purpose of this manuscript was to study grain growth by molecular dynamics simulations in a nanocrystalline polycrystal. For this purpose, we introduced the computational tool GraDe-A for the determination, reconstruction and tracking of grains in FCC polycrystals from MD data sets. This tool was utilized to evaluate the change of properties of the grains. We expect this tool to become a valuable tool for the interpretation of data from MD simulations and the community is invited to contribute to the project \cite{Hoffrogge01}. In addition, recently also a tool capable of discerning orientation from MD datasets was developed \cite{Mahler01}, which in conjunction with the tool introduced here would allow analyzing grains in polycrystals with different crystal structures.
\section{Summary}

A computational tool, which is capable of identifying grain entities by the calculation of per-atom orientation quaternions, was developed. It was demonstrated that the tool is powerful enough to allow for grain-resolved 3D-visualizations of the grains and that it provides useful grain-quantities, such as mean-orientation, volume and rotation angles. A Voronoi-tesselated microstructure with a weak texture was synthesized and used to simulate grain growth at $500$ $\grad$C by molecular dynamics; snapshots from these simulations were used to benchmark the computational tool. The developed algorithm was used to determine global and per-grain kinetics and mean rotation of the grains. 

The results of the simulations substantiate non-linear grain growth kinetics. This behavior was attributed to the dragging of grain boundary dislocations on boundary migration that resulted in decreased kinetics. Additionally, grain rotation was found not to occur significantly and thus grain coalescence occurred only rarely during grain growth in contradiction to the observations by Haslam et al. \cite{Haslam03}. 

Regarding the computational tool GraDe-A, there are various improvements and extensions that are currently under development:

\begin{itemize}
\item{Add support for other lattice-structures}
\item{Identifying and highlighting other network-entities such as faces, edges and junctions}
\item{Calculate grain-properties such as triple line length, turning angle and mean width}
\item{Obtaining topological and connectivity-information for each grain}
\item{Calculating the 5 parameter space of grain boundaries}
\item{Determination of Bernal structures of grain boundaries}
\item{Automatically identify grain-coalescence events}
\end{itemize}

These enhancements would be helpful for the analysis of nanocrystalline grain growth and grain rotation since most of the processes observed are difficult to discuss without knowing exact information on the involved defects and geometries.

\section*{Acknowledgements} 

The authors express their gratitude to the Deutsche Forschungsgemeinschaft (DFG) for financial support (Grants GO 335/44-1) and gratefully acknowledge the computing time granted (JHPC24) by the John von Neumann Institute for Computing (NIC) provided on the supercomputer JURECA at J\"ulich Supercomputing Centre (JSC). P. W. Hoffrogge thankfully acknowledges the computing time granted by the RWTH-Aachen University (RWTH0060) for the execution of his master-thesis project.

\section*{References}
\bibliography{references}

\newpage

\appendix
\newcommand{\In}[1]{\KwIn{\emph{#1}}}
\newcommand{\Out}[1]{\KwOut{\emph{#1}}}
\newcommand{\Result}[1]{\KwResult{\emph{#1}}}
\newcommand{\Data}[1]{\KwData{\emph{#1}}}
\newcommand{\milDir}[1]{\langle #1 \rangle}
\renewcommand{\thealgocf}{\Alph{section}\arabic{algocf}}
\renewcommand{\thefootnote}{\Alph{section}\arabic{footnote}} 
\renewcommand{\theequation}{\Alph{section}.\arabic{equation}}
\section{Appendix}
\subsection{Per-atom orientation calculation}
An orientation is calculated by algorithm \ref{alg:OriCalc} for every atom by taking into account the positions of the nearest neighbors. During the calculation a reduced neighbor vector list $V_R$ is calculated for every atom by algorithm \ref{alg:reducedVects}. This procedure eliminates double occurrences of anti-parallel neighbor unit-vectors by calculating the mean direction $1/2 (v_i-v_j )$ of near antipodal pairs. If a sufficient amount of directions remains, the orientation calculator utilizes them to formulate a transformation matrix $M$, as described in section 2.1. In order to obtain the largest eigenvalue as the resulting quaternion $q$ we use the external open-source linear-algebra solver Armadillo\footnote{http://arma.sourceforge.net}.
\SetAlFnt{\footnotesize}

 \begin{center}
    \scalebox{0.8}{
    \begin{minipage}{1.0\linewidth}
      \begin{algorithm}[H]

\In{The antipodal threshold angle $\Delta\alpha$ and the perpendicular threshold angle $\Delta\beta$ in the current work chosen to be $5\grad$ and $11.5\grad$, respectively.}
\ForEach{atom $a$}{
	Identify all $n_{NN}$ nearest neighbor atoms $A_{NN}$\\
	Calculate the relative positions $V_{NN}$ of $A_{NN}$ to $a$\\
	\tcp{The fcc-lattice has twelve neighbors, too much neighbors may cause problems. }
	\If{$n_{NN} > 12 $}{
		\Return not oriented
	}
	Calculate the unit vectors to $V_{NN}$ and save them in $V_{NN}$\\
	\tcp{Find the antipodal partners and calculate the mean direction}
	Use algorithm \ref{alg:reducedVects} to generate the reduced vector list $V_R$ containing $n_{R}$ vectors\\
	\tcp{For the fcc-lattice at least 6 neighbor directions need to be available}
	\If{$n_{R} < 6$ }{
		\Return not oriented	
	}
	\tcp{Calculate three $\milDir{100}$ directions}
	Empty $\milDir{100}$-vector list $V_{100}$\\
	\For{$i = 0$; $i< n_{R}$; $i$++}{
		$\vec{v_i}$ is the  $i$th vector in $V_{R}$\\
		\If{the number of vectors inside $V_{100}$ is three }{
			break
		}
		\tcp{check all $ij = ji$ combinations}
		\For{$j = i+1$; $j < n_{R}$; $j$++}{
			$\vec{v_j}$ is the  $j$th vector in $V_{R}$\\
			\If{the angle between $\vec{v_i}$ and $\vec{v_j}$ is in the interval $90\grad\pm \Delta\beta$}{
				\tcp{Calculate the crossproduct of  $\vec{v_i}$ with $\vec{v_j}$}
				$\vec{v_{CP}} = \vec{v_i} \times \vec{v_j}$\\
				Append $\vec{v_{CP}}$ to $V_{100}$\\
				\If{the number of vectors inside $V_{100}$ is three}{
					break
				}
			}
			
		}
	}
	\If{the number of vectors inside $V_{100}$ is smaller than three}
	{
		\Return not oriented
	}
	Setup the matrix $M$ as shown in equation (3)\\
	Calculate the orientation quaternion $q$ as a least square solution to $M$, utilizing a $4\times4$-Matrix and computing the corresponding eigenvectors with  Armadillo\\
	\Return the orientation id to $q$
}
\Result{The orientation id to the atoms orientation}
\caption{Algorithm calculating the orientation of an atom in a fcc-lattice structured material by taking into account the near neighborhood.}
\label{alg:OriCalc}

      \end{algorithm}
    \end{minipage}%
    }
  \end{center}

\clearpage

 \begin{center}
    \scalebox{0.85}{
    \begin{minipage}{1.0\linewidth}
      \begin{algorithm}[H]
			
\In{
	\\Vector list $V_{NN}$ containing $n_{NN}$ nearest neighbor unit-vectors of an atom\\
	The antipodal threshold angle $\Delta\alpha$ 
	}
	Empty reduced-vector list $V_{R}$\\
		\For{$i = 0$; $i< n_{NN}$; $i$++}{
			$v_i$ is the  $i$th vector in $V_{NN}$\\
			\tcp{check all $ij = ji$ combinations}
			\For{$j = i+1$; $j < n_{NN}$; $j$++}{
				$v_j$ is the  $j$th vector in $V_{NN}$\\
				\If{both $v_i$ and $v_j$ are unpaired}{
					\If{the angle between $v_i$ and $v_j$ is in the interval $180\grad\pm \Delta\alpha$}{
						Append $\frac{1}{2}(v_i - v_j)$ to $V_R$\\
						$v_i$ and $v_j$ are marked as paired
					}
				}
			}
			\tcp{Poor $v_i$ didn't find a partner ?}
			\If{$v_i$ is still unpaired}{
				Append $v_i$ to $V_R$	
			}
		}
\Result{The reduced vector list $V_R$}
\caption{Algorithm reducing a nearest neigbor vector list by eliminating twice occurrences of antipodal vectors.}
\label{alg:reducedVects}
			
      \end{algorithm}
    \end{minipage}%
    }
  \end{center}



\newpage
\subsection{Grain identification and reconstruction}
In order to group atoms together into a grain entity, two different criteria were introduced. A \emph{local criterion} that compares the orientation of a candidate atom i.e. that of the atom to be assigned to a certain grain and that of the reference atom, referred to as parent atom. A \emph{global criterion} compares the average orientation of the grain and the orientation of the candidate atom.

\newpage

 \begin{center}
    \scalebox{0.85}{
    \begin{minipage}{1.0\linewidth}
      \begin{algorithm}[H]
			
\renewcommand{\;}{\\}
	\Data{Position data for all atoms}
	\Result{A certain number of grains}
	\ForEach{atom $a$}{
		\If{$a$ has been assigned to any grain}{ continue\;}
		new grain $G$\;
		empty atom list $l_N$\;
		empty atom list $l_{NN}$\;
		add $a$ to $l_N$ with $a$ as parent atom\;
		\While{number of entries of $l_N > 0$}{
			\ForEach{atom $a_n$ in $l_N$}{
				\If{$a_n$ has been assigned to any grain}{continue\;}
				\tcp{local criterion}
				\If{$a_n$ and its parent atom have not a very close orientation}{continue\;}
				\tcp{global criterion}
				\If{the orientation of $a_n$ and the average orientation of $G$ are not close}{continue\;}
				assign $a_n$ to $G$\;
				add all nearest neighbors of $a_n$ to $l_{NN}$ with $a_n$ as parent atom\;
			}
			$l_N = l_{NN}$\;
			$l_{NN} = $ new empty list\;
		}
		\If{$G$ does not comprise a sufficient amount of atoms}{
			dissociate all atoms of $G$\;
			delete $G$\;
		}
	}
	\caption{Grain-identification algorithm grouping areas of similarly oriented atoms into grain entities.}
	\label{alg:GrainIdentification}
			
      \end{algorithm}
    \end{minipage}%
    }
  \end{center}

	

\newpage
\subsection{Orphan atom adoption}
The atoms that were not successfully assigned to a grain by the previous algorithm are referred to as \emph{orphan atoms}. The current algorithm is used to assign these atoms to a grain in a procedure deemed as atom adoption, which is based on the most frequent occurrences of the same grain. A threshold for the minimum number of occurrences of a grain was implemented to suppress a random pick behavior. The procedure is repeated for every atom until no further adoptions can occur in the dataset.

\newpage

 \begin{center}
    \scalebox{0.85}{
    \begin{minipage}{1.0\linewidth}
      \begin{algorithm}[H]
			
	\renewcommand{\;}{\\}
	\Data{Position data for all atoms}
	\Result{A certain number of grains}
	\ForEach{orphan atom $a_{orph}$}{
		count(all grains) = 0\;
		\If{$a_{orph}$ has been adopted by any grain}{ continue\;}
		\ForEach{neighbor atom $a_n$ to $a_{orph}$}{
			\If{$a_n$ is assigned to any grain}{
				count(grain($a_n$)) ++\;
			}
		}
		identify the grain $g_{MF}$ with the highest count $c_{MF}$\;
		\If{$c_{MF}$ $\geq$ minimum number of grain occurrence}{
			adopt $a_n$ by $g_{MF}$\;
			\tcp{$a_n$ is now assigned to $g_{MF}$}
		}
	}
	\caption{Algorithm assigning atoms to preexisting grains by taking into account the grain-membership of nearest neighbor atoms.}
	\label{alg:OrphanAtomAdopt}

      \end{algorithm}
    \end{minipage}%
    }
  \end{center}



\newpage
\subsection{Grain tracking over time}

Algorithm \ref{alg:TimeDependentGrainDetect} finds a corresponding grain-entity from a previously analyzed data set. Both the center-of-mass (COM) and the orientation of the grains are utilized as criteria to match couples from the data sets.

 \begin{center}
    \scalebox{0.85}{
    \begin{minipage}{1.0\linewidth}
      \begin{algorithm}[H]
			
	\renewcommand{\;}{\\}
	\In{a newly identified grain $g$}
	\Data{\\
		Volume $V_G$ of the grain $g$\\
		some factor $a$
	}
	mapped grain $g_{mapped}$
	grain id $id_{G} = $ \emph{NO\_GRAIN}\;
	$d_{max} = a\left(\frac{V_{G}}{4/3 \pi}\right)^{1/3}$\;
	$d_{min} = d_{max}$\;
	\ForEach{Grain  $g_{prev}$ from the previous timestep}{
		\If{$g_{prev}$ is already assigned to any grain}{continue\;}
		\tcp{Center-of-mass distance-criterion}
		$d = $ distance between COM($g$) and COM($g_{prev}$)\;
		\If{$d > d_{max}$}{continue\;}
		\tcp{Misorientation-criterion}
		$\Delta\varphi = $ misorientation between the average orientations of $g$ and $g_{prev}$\;
		\If{$\Delta\varphi > \Delta\varphi_{max}$}{continue\;}
		\tcp{All criteria fulfilled}
		\If{$d \leq d_{min}$ }{
			$d_{min} = d$\;
			$g_{mapped} = g_{prev}$\;
			$id_{G} = id(g_{prev})$\;
		}
	}
	\If{$id_G \neq $ \emph{NO\_GRAIN}}{
		\tcp{do not allow a double assignment}
		set $g_{mapped}$ as assigned\;
		\Return $id_G$\;
	}
	\Return a new id
	\caption{Algorithm assigning a currently detected grain to a grain-entity from a previous timestep.}
	\label{alg:TimeDependentGrainDetect}
	
      \end{algorithm}
    \end{minipage}%
    }
  \end{center}

\newpage
\subsection{Rotation quaternions and unique orientations}
During the determination of the per-atom orientation, owing to the symmetry of the cubic lattice different but equivalent orientation quaternions may be calculated. For this reason, it is necessary to reduce the orientation to a unique one. This done by algorithm \ref{alg:CubUniqueOrients} by the utilization of Eqs. (\ref{eqn:CubicEquivQuatsFirst})-(\ref{eqn:CubicEquivQuatsLast}):

 \begin{center}
    \scalebox{0.85}{
    \begin{minipage}{1.0\linewidth}
      \begin{algorithm}[H]
			
	\In{An orientation quaternion $q$}
	Cosine Angle $C_{max} = 0$\\
	The unique quaternion to identify $q_u$\\
	\ForEach{24 cubically equivalent quaternions $q_i$ to $q$}
	{
		\If{$|q_i[0]| > C_{max}$} {
			$C_{max} = |q_i[0]|$\\
			$q_u = q_i$
		}
	}
	\If{$q_u[0] < 0 $}{$q_u = -q_u$}
	\Return{$q_u$}
	\caption{Algorithm to obtain the unique cubical orientation quaternion $q_{uo}$ to a quaternion $q$.}
	\label{alg:CubUniqueOrients}
      \end{algorithm}
    \end{minipage}%
    }
  \end{center}
	
\subsection*{List of the 24 cubically-equivalent orientation quaternions:}

\fontsize{10}{12}\selectfont

\begin{align}
q_1^T=[a_0;a_1,a_2,a_3]%
\label{eqn:CubicEquivQuatsFirst}%
\end{align}
\begin{align}
q_2^T=[-a_1;a_0,a_3,-a_2]\\
q_3^T=[-a_2;-a_3,a_0,a_1]\\
q_4^T=[-a_3;a_2,-a_1, a_0]
\end{align}
\begin{align}
q_5^T&=0.5[a_0 - a_1 - a_2 - a_3; a_0 + a_1 + a_2 - a_3, a_0 - a_1 + a_2 + a_3, a_0 + a_1 - a_2 + a_3]\\
q_6^T&=0.5[ a_0 +  a_1 +  a_2 +  a_3; - a_0 +  a_1 -  a_2 +  a_3, - a_0 +  a_1 +  a_2 -  a_3, - a_0 -  a_1 +  a_2 +  a_3]\\
q_7^T&=0.5[ a_0 -  a_1 +  a_2 -  a_3;  a_0 +  a_1 +  a_2 +  a_3, - a_0 -  a_1 +  a_2 +  a_3,  a_0 -  a_1 -  a_2 +  a_3]\\
q_8^T&=0.5[ a_0 +  a_1 -  a_2 +  a_3; - a_0 +  a_1 -  a_2 -  a_3,  a_0 +  a_1 +  a_2 -  a_3, - a_0 +  a_1 +  a_2 +  a_3]\\
q_9^T&=0.5[ a_0 +  a_1 -  a_2 -  a_3; - a_0 +  a_1 +  a_2 -  a_3,  a_0 -  a_1 +  a_2 -  a_3,  a_0 +  a_1 +  a_2 +  a_3]\\
q_{10}^T&=0.5[ a_0 -  a_1 +  a_2 +  a_3;  a_0 +  a_1 -  a_2 +  a_3, - a_0 +  a_1 +  a_2 +  a_3, - a_0 -  a_1 -  a_2 +  a_3]\\
q_{11}^T&=0.5[ a_0 +  a_1 +  a_2 -  a_3; - a_0 +  a_1 +  a_2 +  a_3, - a_0 -  a_1 +  a_2 -  a_3,  a_0 -  a_1 +  a_2 +  a_3]\\
q_{12}^T&=0.5[ a_0 -  a_1 -  a_2 +  a_3;  a_0 +  a_1 -  a_2 -  a_3, a_0 +  a_1 +  a_2 +  a_3, - a_0 +  a_1 -  a_2 +  a_3]
\end{align}
\begin{align}
q_{13}^T=\frac{1}{\sqrt{2}}[a_0 - a_1; a_0 + a_1, a_2 + a_3, -a_2 + a_3]\\
q_{14}^T=\frac{1}{\sqrt{2}}[a_0 - a_2; a_1 - a_3, a_0 + a_2, a_1 + a_3]\\
q_{15}^T=\frac{1}{\sqrt{2}}[a_0 - a_3; a_1 + a_2, -a_1 + a_2, a_0 + a_3]\\
q_{16}^T=\frac{1}{\sqrt{2}}[-a_1 - a_2; a_0 - a_3, a_0 + a_3, a_1 - a_2]
\end{align}
\begin{align}
q_{17}^T=\frac{1}{\sqrt{2}}[-a_2 - a_3; a_2 - a_3, a_0 - a_1, a_0 + a_1]\\
q_{18}^T=\frac{1}{\sqrt{2}}[-a_1 - a_3; a_0 + a_2, -a_1 + a_3, a_0 - a_2]\\
q_{19}^T=\frac{1}{\sqrt{2}}[a_0 + a_1; -a_0 + a_1, a_2 - a_3, a_2 + a_3]\\
q_{20}^T=\frac{1}{\sqrt{2}}[a_0 + a_2; a_1 + a_3, -a_0 + a_2, -a_1 + a_3]\\
q_{21}^T=\frac{1}{\sqrt{2}}[a_0 + a_3; a_1 - a_2, a_1 + a_2, -a_0 + a_3]\\
q_{22}^T=\frac{1}{\sqrt{2}}[a_1 - a_2; -a_0 - a_3, a_0 - a_3, a_1 + a_2]\\
q_{23}^T=\frac{1}{\sqrt{2}}[a_2 - a_3; a_2 + a_3, -a_0 - a_1, a_0 - a_1]\\
q_{24}^T=\frac{1}{\sqrt{2}}[a_1 - a_3; -a_0 + a_2, -a_1 - a_3, a_0 + a_2]%
\label{eqn:CubicEquivQuatsLast}%
\end{align}

\end{document}